\documentclass[a4paper,10pt,twoside]{article}

\usepackage{a4wide}  
\usepackage{geometry}
\usepackage{amsmath}
\usepackage{amssymb}
\usepackage[greek, british]{babel}      
\usepackage{fancyhdr}                   
\usepackage{graphicx}                  
\usepackage{grffile}                 
\usepackage{pifont}                     
\usepackage{paralist}                   
\usepackage[usenames,dvipsnames]{color} 
\usepackage{moreverb}                 
\usepackage{lscape}                   
\usepackage{tocbibind}
\usepackage{verbatim}                
\usepackage{datetime}                
\usepackage{array}                     
\usepackage{color}
\definecolor{gray}{rgb}{0.35,0.35,0.35}
\usepackage[pagebackref=true,
            colorlinks=true,
            linkcolor=gray,
            bookmarks=true,
            filecolor=gray,
            urlcolor=gray,
            citecolor=gray
           ]{hyperref} 
\usepackage{lineno}                   
\usepackage{longtable}                
\usepackage[section]{placeins}       
\usepackage{url}                      
\usepackage{makeidx}                
\usepackage{hyperref}                 
\usepackage{subfig}                 
\usepackage{booktabs}                
\usepackage{rotating}       
\renewcommand{\footrulewidth}{0.4pt} 
\fancypagestyle{plain}{
\fancyhead[LE,RO]{}
\fancyhead[LO,RE]{}
\renewcommand{\headrulewidth}{0pt} 
}
{
\fancyhead[LE,RO]{\slshape \leftmark} 
\fancyhead[LO,RE]{\slshape } 
}

\usepackage[format=plain,
            labelfont={bf,it},
            textfont=it]{caption}
\usepackage{pdfpages}
\usepackage{pdflscape}
\usepackage{graphicx} 
\usepackage{float}
\usepackage{amsmath,verbatimbox}
\usepackage{verbatimbox}

\usepackage{amsmath}
\usepackage{amssymb}
\usepackage{textcomp, gensymb}
\usepackage{gensymb}
\usepackage{setspace}
\usepackage{mathtools}
\usepackage{optidef}
\usepackage{enumitem}   
\usepackage{arydshln}

\begin{document}

\renewcommand{\footrulewidth}{0pt}
\renewcommand{\headrulewidth}{0pt}
\renewcommand{\figurename}{Fig.}
\renewcommand*{\figureautorefname}{Fig.}

\newcommand{\sms}{\kern 0.16667em}
\newcommand{\GammaBarTwo}{$\bar{\gamma}_{2D}^{s}$}
\newcommand{\GammaHatTwo}{$\hat{\gamma}_{2D}^{s}$}
\newcommand{\GammaBar}{$\bar{\gamma}^{s}$}
\newcommand{\GammaBarO}{$\bar{\gamma}^{o}$}
\newcommand{\GammaBarCombi}{$\bar{\gamma}^{s/o}$}
\newcommand{\GammaBarTwoCombi}{$\bar{\gamma}^{s/o}_{2D}$}
\newcommand{\GammaHat}{$\hat{\gamma}^{s}$}
\newcommand{\SvecHatTwo}{$\hat{\vec{s}}_{\sms 2D}$}
\newcommand{\SvecBarTwo}{$\bar{\vec{s}}_{\sms 2D}$}
\newcommand{\OvecHatTwo}{$\hat{\vec{o}}_{\sms 2D}$}
\newcommand{\OvecBarTwo}{$\bar{\vec{o}}_{\sms 2D}$}
\newcommand{\NvecBarTwo}{$\bar{\vec{n}}_{\sms 2D}$}
\newcommand{\TvecBarTwo}{$\bar{\vec{t}}_{\sms 2D}$}
\newcommand{\SvecHat}{$\hat{\vec{s}}$}
\newcommand{\SvecBar}{$\bar{\vec{s}}$}
\newcommand{\OvecHat}{$\hat{\vec{o}}$}
\newcommand{\OvecBar}{$\bar{\vec{o}}$}
\newcommand{\NvecBar}{$\bar{\vec{n}}$}
\newcommand{\TvecBar}{$\bar{\vec{t}}$}
\newcommand{\NvecHat}{$\hat{\vec{n}}$}
\newcommand{\TvecHat}{$\hat{\vec{t}}$}
\newcommand{\RHat}{$\hat{R}$}
\newcommand{\RBar}{$\bar{R}$}

\setcounter{page}{0}
\pagenumbering{arabic}

\Large
\begin{center}
\noindent \textbf{SLIDE: Automated Identification and Quantification of Grain Boundary Sliding and Opening in 3D}\\
\end{center}

\begin{center}
\large
\noindent C.J.A. Mornout\textsuperscript{a}, G. Slokker\textsuperscript{a}, T. Vermeij\textsuperscript{a,b}, D. König\textsuperscript{a}, J.P.M. Hoefnagels\textsuperscript{a,*} 
\end{center}

\footnotesize
\noindent \textsuperscript{a}\textit{Dept. of Mechanical Engineering, Eindhoven University of Technology, 5600MB Eindhoven, The Netherlands} \\
\noindent \textsuperscript{b}\textit{Laboratory for Mechanics of Materials and Nanostructures, Swiss Federal Laboratories for Materials Science and
Technology (EMPA), Feuerwerkerstrasse 39, 3602 Thun, Switzerland}

\vspace{\baselineskip}

\noindent \textsuperscript{*}Corresponding author.
\textit{Email address:} j.p.m.hoefnagels@tue.nl (J.P.M. Hoefnagels)\\
\textit{First author email address:} c.j.a.mornout@tue.nl (C.J.A. Mornout)
\normalsize

\section*{Abstract}

Grain Boundary (GB) deformation mechanisms such as Sliding (GBS) and Opening (GBO) are prevalent in alloys at high homologous temperatures but are hard to capture quantitatively. We propose an automated procedure to quantify 3D GB deformations at the nanoscale, using a combination of precisely aligned Digital Image Correlation (DIC), electron backscatter diffraction, optical profilometry, and in-beam secondary electron maps. The framework, named \textbf{S}liding identification by \textbf{L}ocal \textbf{I}ntegration of \textbf{D}isplacements across \textbf{E}dges (\textbf{SLIDE}), (i) distinguishes GBS from GBO, (ii) computes the datapoint-wise measured in-plane displacement gradient tensor (from DIC), (iii) projects this data onto the theoretical GBS tensor to reject near-GB plasticity/elasticity/noise, and (iv) adds the out-of-plane step from optical profilometry to yield the local 3D GBS/GBO vector; automatically repeated for each $\sim$50nm-long GB segment. SLIDE is validated on a virtual experiment of discrete 3D sliding, and successfully applied to Zn-coated steel experiments, yielding quantitative GBS/GBO activity maps.

\vspace{\baselineskip}
\small
\textit{Keywords:} grain boundary sliding, electron backscattering diffraction (EBSD), scanning electron microscopy (SEM), digital image correlation (DIC), optical profilometry
\normalsize

\vfill
\noindent\makebox[\textwidth]{%
  \makebox[0pt][l]{\footnotesize \textit{Preprint submitted to Arxiv}}%
  \hfill
  \makebox[0pt][r]{\footnotesize \textit{July 23, 2025}}%
}


\newpage


\section*{Graphical Abstract}
\begin{figure}[ht!]
    \centering
    \includegraphics[width=\linewidth]{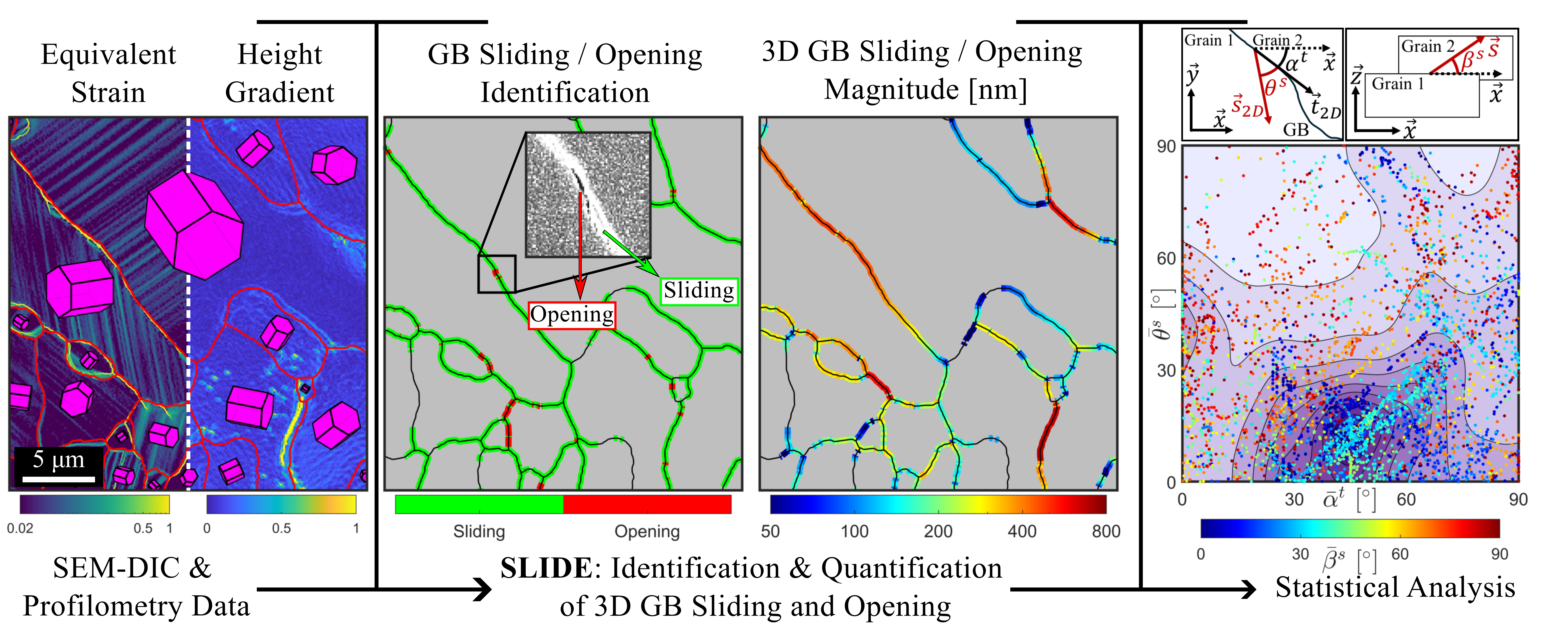}
\end{figure} 

\section*{Highlights}
\begin{itemize}
    \item Automatic 3D Sliding/Opening vector from aligned SEM-DIC \& profilometry GB maps. 
    \item  Projection on theoretical GB sliding tensor rejects near-GB  crystal plasticity/elasticity/noise.
    \item Sliding/Opening robustly distinguished using In-Beam SE aligned to EBSD.
    \item 3D Sliding/Opening maps with $\sim$50nm resolution enable large statistical GB analysis.
\end{itemize}

\newpage

Deformation mechanisms on Grain Boundaries (GBs) have been investigated for decades and are commonly divided between GB Sliding (GBS) \cite{li_stress-induced_1953}, GB cracking or Opening (GBO) \cite{vincent_relationship_2006,bhattacharya_transgranular_2020}, and GB migration \cite{gleiter_mechanism_1969,gottstein_grain_1998}. GB deformation is frequently observed in fine-grained materials \cite{zhao_role_2018,wang_grain_2014,mungole_contribution_2015} or materials at high homologous temperature \cite{texier_strain_2024,kang_grain_2020,langdon_role_1993}; observations of GBS at room temperature are therefore common for metals and alloys with a low melting temperature such as Mg \cite{orozco-caballero_how_2017,yavuzyegit_mapping_2023} and Zn \cite{bell_grain-boundary_1968,sheikh-ali_dislocation_1990,takahashi_coupling_1985}.

Many experimental studies have been devoted to observing and quantifying GBS for bicrystals \cite{sheikh-ali_dislocation_1990,takahashi_coupling_1985,wei_direct_2021} and polycrystals \cite{texier_strain_2024,kang_grain_2020,orozco-caballero_how_2017,sangid_deformation_2019,ando_internal_2016}. Recent works analyzed GBS by combining Scanning Electron Microscopy-based Digital Image Correlation (SEM-DIC) with Electron Backscatter Diffraction (EBSD) data \cite{yavuzyegit_mapping_2023,linne_effect_2020,dimanov_deformation_2021,jullien_grain_2024}. For example, Yavuzyegit \textit{et al.} analyzed in-plane GBS by calculating the tangential displacements along GBs using an algorithm that detects GBs in a shear strain map \cite{yavuzyegit_mapping_2023} and Linne \textit{et al.} quantified in-plane GB displacements and analyzed their interaction with intragranular slip \cite{linne_effect_2020}. 

Measurements of out-of-plane displacements, using atomic force microscopy or Optical surface Profilometry (OP), can further help identification of plasticity, as shown for intragranular slip \cite{liu_-plane_2019,yin_three-dimensional_2023}. Very recently, Jullien \textit{et al.} employed OP to measure the out-of-plane step of each GB, obtaining a 3D displacement vector \cite{jullien_grain_2024}. A step in the right direction, however, GBO was not distinguished from GBS, which is a crucial step since GBO can be a critical damage mechanism \cite{vincent_relationship_2006,jullien_grain_2024,song_relation_2012,parisot_deformation_2004}. Moreover, the out-of-plane step was assumed to be constant over the whole $\mu m$-sized GB from triple point to triple point. As we will show, the 3D GB displacement vector varies locally between triple points, sometimes even alternating between GBS and GBO in the same GB. More fundamentally, in all aforementioned works it is assumed that the measured in-plane GB displacement, obtained by comparing the displacement from DIC on both sides, can completely be attributed to GBS. This was never validated by comparing to the solid mechanics description of the local kinematics of GBS. Indeed, our fundamental analysis reveals that the locally measured displacements at the GB are often incompatible with the analytical kinematic description of GBS, caused by factors such as intragranular slip near the GB, elastic rotation/deformation, measurement noise, etc. 

Therefore, a general and automated method to locally identify and quantify 3D GBS/GBO at the nanoscale is desired. The method should, for each $\sim$50nm GB segment, 1) distinguish between GBS/GBO, 2) accurately compute a 3D GB displacement vector, 3) project the displacement onto the kinematic description of GBS/GBO in order to prevent that non-GB-based deformations contribute to GBS direction/magnitude, and 4) be fully automated to enable statistical analysis of large polycrystalline datasets.  

In this work, we propose a framework to fully identify and quantify the 3D GB deformation and distinguish between GBS/GBO, by combining EBSD-based orientation maps, GBO maps from In-Beam Secondary Electron (IBSE) imaging, SEM-DIC-based nanoscale strain maps, and out-of-plane displacement maps from OP, all aligned with the recently introduced nanomechanical alignment framework published in \cite{vermeij_nanomechanical_2022}. The novel approach, termed \textbf{S}liding identification by \textbf{L}ocal \textbf{I}ntegration of \textbf{D}isplacements across \textbf{E}dges (\textbf{SLIDE}), is inspired by the recently proposed framework for 'Slip Systems based Local Identification of Plasticity' (SSLIP) \cite{vermeij_automated_2023}. SSLIP uses the measured in-plane displacement gradient tensor and EBSD data to calculate, for each data point, the most likely (combination of) 3D slip system activity, marking a significant step forward in intragranular slip identification \cite{vermeij_martensite_2023,vermeij_quasi-2d_2024,konig_direct_2025,vermeij_sslip_2025}. In similar fashion to SSLIP, the SLIDE framework yields complete, local, and automated identification of nanoscale plasticity on the GB, additionally identifying GBO and out-of-plane displacements. The SLIDE code will be freely available on Github.

SLIDE starts from the solid mechanics description of the 3D kinematics of a sliding/opening motion on a GB segment with normal $\vec{n}$ and tangential $\vec{t}$, described by the 3D displacement gradient tensors $\mathbf{H}^{GBS}$/$\mathbf{H}^{GBO}$ as:
\begin{subequations}
\begin{eqnarray}
 \mathbf{H}^{GBS} = \gamma^s \ \vec{s} \ (\alpha^s,\beta^s) \otimes \vec{n} \ (\alpha^n,\beta^n), \  \textrm{where} \ \vec{s} \cdot \vec{n} = 0, \\
 \mathbf{H}^{GBO} = \gamma^o \ \vec{o} \  (\alpha^o,\beta^o) \otimes \vec{n} \ (\alpha^n,\beta^n), \  \textrm{where} \ \vec{o} \cdot \vec{n} \neq 0,
\end{eqnarray}
\end{subequations}

\noindent where $\vec{s}$/$\vec{o}$ are the sliding/opening vectors with magnitude $\gamma$, and $\otimes$ denotes a dyadic product. Each vector is described by the in-plane angle $\alpha$ and the out-of-plane angle $\beta$ between the vector and the direction of tension $\vec{x}$. From in-plane SEM-DIC data, four components of the experimental displacement gradient tensor $\mathbf{H}^{exp}_{2D}$ can be computed:
\begin{equation}
\mathbf{H}^{exp}_{2D}=\vec{\nabla}_0 \vec{u}=\left[\begin{array}{cc}
H_{x x}^{e x p} & H_{x y}^{e x p} \\
H_{y x}^{e x p} & H_{y y}^{e x p}
\end{array}\right],
\label{eq:Hexp}
\end{equation}

\noindent where $\vec{\nabla}_0 \vec{u}$ is the gradient of the local displacement vector $\vec{u}$. $\mathbf{H}^{exp}_{2D}$ is compared to the in-plane part (indicated by the dashed squares) of the 3D $\mathbf{H}^{GBS}$ and $\mathbf{H}^{GBO}$ tensors:
\begin{subequations}
\begin{eqnarray}
\mathbf{H}^{GBS}= \gamma^s \left[\begin{array}{cc} \begin{array}{:cc:}
\hdashline \vec{s}{_x}\vec{n}{_x} \ (\alpha^s,\alpha^n) & \vec{s}{_x}\vec{n}{_y} \ (\alpha^s,\alpha^n) \\ \vec{s}{_y}\vec{n}{_x} \ (\alpha^s,\alpha^n) & \vec{s}{_y}\vec{n}{_y} \ (\alpha^s,\alpha^n) \\ \hdashline \end{array} \begin{array}{cc} \vec{s}{_x}\vec{n}{_z} \ (\alpha^s,\beta^n)  \\ \vec{s}{_y}\vec{n}{_z} \ (\alpha^s,\beta^n) \end{array} \\ \begin{array}{cc} \vec{s}{_z}\vec{n}{_x} \ (\beta^s,\alpha^n) &  \vec{s}{_z}\vec{n}{_y} \ (\beta^s, \alpha^n) \end{array} \begin{array}{cc} \vec{s}{_z}\vec{n}{_z} \ (\beta^s,\beta^n) \end{array} \end{array}\right], \\
\mathbf{H}^{GBO}= \gamma^o \left[\begin{array}{cc} \begin{array}{:cc:} \hdashline \vec{o}{_x}\vec{n}{_x} \ (\alpha^o,\alpha^n) & \vec{o}{_x}\vec{n}{_y} \ (\alpha^o,\alpha^n) \\ \vec{o}{_y}\vec{n}{_x} \ (\alpha^o,\alpha^n) & \vec{o}{_y}\vec{n}{_y} \ (\alpha^o,\alpha^n) \\ \hdashline \end{array} \begin{array}{cc} \vec{o}{_x}\vec{n}{_z} \ (\alpha^o,\beta^n)  \\ \vec{o}{_y}\vec{n}{_z} \ (\alpha^o,\beta^n) \end{array} \\ \begin{array}{cc} \vec{o}{_z}\vec{n}{_x} \ (\beta^o,\alpha^n) &  \vec{o}{_z}\vec{n}{_y} \ (\beta^o, \alpha^n) \end{array} \begin{array}{cc} \vec{o}{_z}\vec{n}{_z} \ (\beta^o,\beta^n) \end{array} \end{array}\right]. \label{eq:GBS_DOF2}
\end{eqnarray}
\label{eq:GBS_DOF}
\end{subequations}

\noindent The in-plane part of GBS/GBO is described by three independent Degrees of Freedom (DOFs): $\gamma^{s}$/$\gamma^{o}$, $\alpha^{s}$/$\alpha^{o}$ and $\alpha^n$ which is known from the GB geometry. Therefore, for each datapoint, SLIDE matches $\mathbf{H}^{exp}_{2D}$ to $\mathbf{H}^{GBS}_{2D}$ or $\mathbf{H}^{GBO}_{2D}$ by finding the two optimized parameters $\gamma_{opt}$ and $\alpha_{opt}$:
\begin{equation}
\mathbf{H}^{exp}_{2D} \approx
\left\{
\begin{array}{ll}
    \mathbf{H}^{GBS}_{2D} \ (\gamma_{opt}^{s}, \alpha_{opt}^{s}) & \text{for GBS} \\
    \mathbf{H}^{GBO}_{2D} \ (\gamma_{opt}^{o}, \alpha_{opt}^{o}) & \text{for GBO}
\end{array}
\right.
\end{equation}

\noindent The optimized DOFs are found by minimizing the $\mathcal{L}_2$ residual norm $R$ of the normalized difference between $\mathbf{H}^{exp}_{2D}$ and $\mathbf{H}^{GBS}_{2D}$ or $\mathbf{H}^{GBO}_{2D}$, which for GBS equates to:
\begin{equation}
(\gamma_{opt}^{s}, \alpha_{opt}^{s}) = \operatorname*{Argmin}_{(\gamma^{s}, \alpha^{s})} \{ R \} = \operatorname*{Argmin}_{(\gamma^{s}, \alpha^{s})} \{ \frac{\| \mathbf{H}^{exp}_{2D} -\mathbf{H}^{GBS}_{2D} \ (\gamma^{s}, \alpha^{s}) \|^{2D}}{ \| \mathbf{H}^{exp}_{2D} \|^{2D}} \}.
\label{eq:argmin}
\end{equation}

\noindent The solution procedure for GBO is the same. Note that, in contrast to this fundamental description, experimental data is obtained in a discrete datapoint-wise fashion and with limited spatial resolution. In the following, any quantity that is applicable to a field datapoint is denoted by $\hat{\Box}$, while $\bar{\Box}$ denotes the (underlying) step in a quantity's value at the GB. To obtain the sliding vector \SvecBarTwo\ at each GB segment from the datapoint-wise field solutions for \SvecHatTwo$(\hat{\gamma}_{opt}^{s}, \hat{\alpha}_{opt}^{s})$ obtained by solving \autoref{eq:argmin}, integration of \SvecHatTwo\ along \NvecBarTwo\ is required, which is illustrated in a virtual experiment. 

\begin{figure}[ht!]
    \centering
    \includegraphics[width=\linewidth]{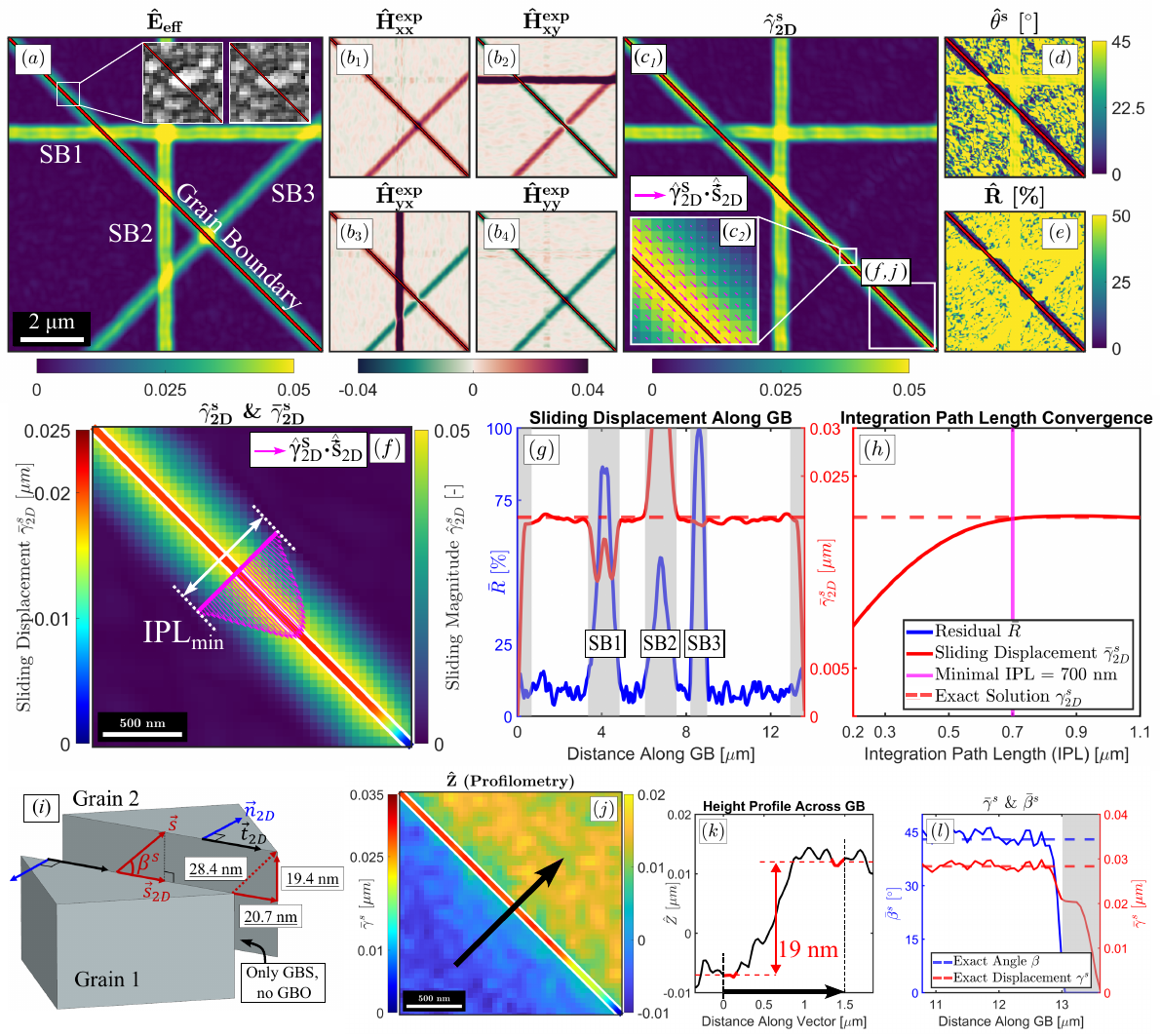}
    \caption{\textbf{Overview of the virtual experiment for 3D GBS.} The discrete in-plane and out-of-plane sliding steps are schematically visualized in (i). (a) Effective strain showing in-plane GBS and three slip bands (SB); $E_{eff}= \sqrt{0.5\left(H_{xx}^{exp}-H_{yy}^{exp}\right)^2+0.5\left(H_{xy}^{exp}+H_{yx}^{exp}\right)^2}$ \cite{dieter_mechanical_1988}. The insert in (a) shows the sliding step and the effect of added noise in the deformed image. (b\textsubscript{1-4}) In-plane $\hat{\mathbf{H}}^{exp}$ components from DIC, using \autoref{eq:Hexp}. (c\textsubscript{1}) Sliding magnitude \GammaHatTwo\ for each field datapoint. (c\textsubscript{2}) GBS vectors scaled by their magnitude, i.e., \GammaHatTwo$\cdot$\SvecHatTwo. (d) Angle $\hat{\theta}^{s}$ between \SvecHatTwo\ and \TvecBarTwo. (e) Normalized residual $\hat{R}$. (f) Close-up of sliding magnitude field \GammaHatTwo, and GB-integrated in-plane sliding displacement \GammaBarTwo\ overlaid on the GB. (g) GB-integrated sliding displacement \GammaBarTwo\ and residual $\bar{R}$. (h) Average sliding displacement \GammaBarTwo\ for different integration path lengths (SB locations excluded). (j) Out-of-plane step of 19.4nm of calibration specimen measured through profilometry, including \GammaBar\ overlaid on the GB. (k) Calculation of height step for the GB segment shown by a black arrow in (j), where the bold red sections (200nm long) of the height profile are used to calculate mean heights (red dashed lines)on both sides of the GB. (l) 3D sliding displacement \GammaBar\ and out-of-plane angle $\bar{\beta}^{s}$ along the GB.}
    \label{fig:Figure1}
\end{figure} 

\begin{figure}[ht!]
    \centering
    \includegraphics[width=\linewidth]{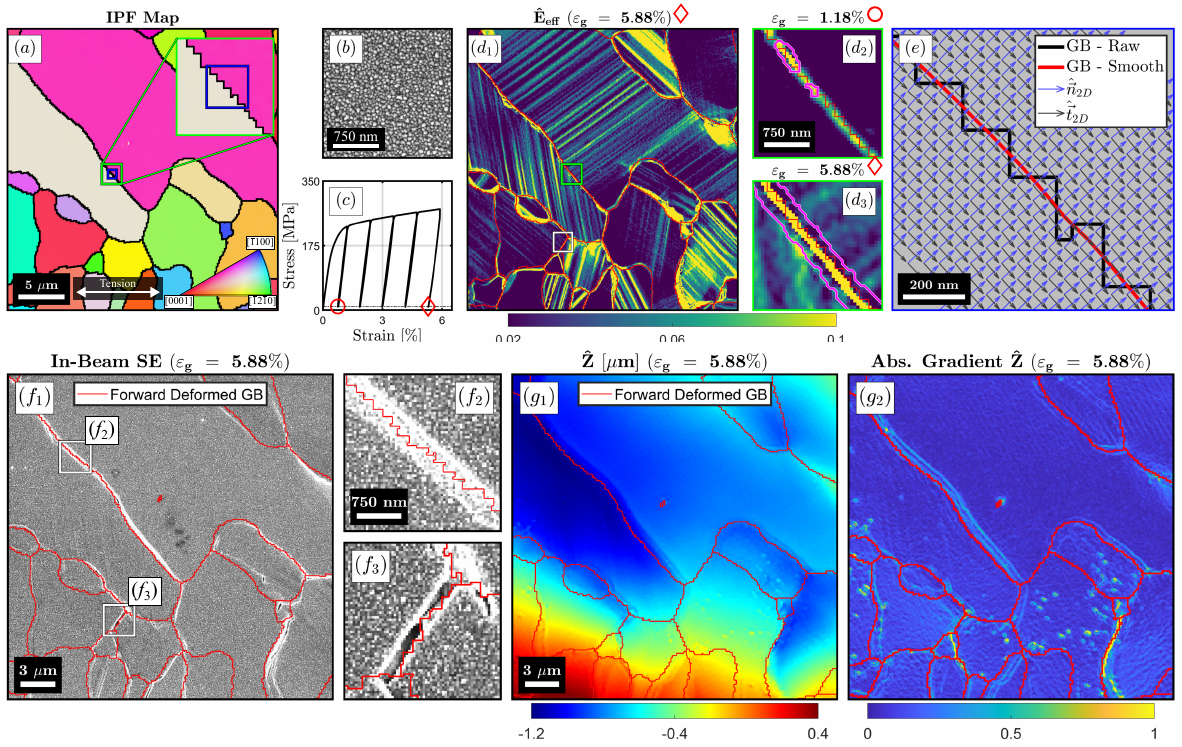}
    \caption{\textbf{Overview of experimental data for one region of interest.} (a) Inverse Pole Figure (Z-IPF) map, in which the direction of tension is indicated. (b) Close-up of InSn DIC speckle pattern \cite{hoefnagels_one-step_2019}, also used for the virtual experiment in \autoref{fig:Figure1}. (c) Engineering stress-strain curve, red circle and diamond correspond to increments shown in (d\textsubscript{1-3}). (d\textsubscript{1}) $\hat{E}_{eff}$ map. The green and white squares highlight regions of GBS and GBO, shown in \autoref{fig:Figure3} and \autoref{fig:Figure4}. (d\textsubscript{2,3}) Close-ups of $\hat{E}_{eff}$ for the green area shown in (a,d\textsubscript{1}). Regions that required DIC displacement interpolation \cite{vermeij_martensite_2023} are encircled in magenta. (e) Close-up of the blue area in (a), displaying the raw and smooth GB, including a band of $\hat{\vec{n}}_{\sms 2D}$ and $\hat{\vec{t}}_{\sms 2D}$ vectors assigned to neighboring datapoints. (f\textsubscript{1},g\textsubscript{1}) IBSE image and height map at the final deformation increment, including forward-deformed GBs. (f\textsubscript{2,3}) Close-ups of (f\textsubscript{1}), showing GBS (f\textsubscript{2}) and GBO (f\textsubscript{3}). (g\textsubscript{2}) Absolute gradient of the height map ($\sqrt{\left(\frac{d z}{d x}\right)^2+\left(\frac{d z}{d y}\right)^2}$), highlighting locations of strong out-of-plane GB plasticity.}
    \label{fig:Figure2}
\end{figure}

A virtual case of 3D GBS, composed of a discrete in-plane sliding step along a diagonal GB, and an out-of-plane step, is shown in \autoref{fig:Figure1}(i). We start with a real IBSE image of a InSn DIC speckle pattern \cite{hoefnagels_one-step_2019}, shown in \autoref{fig:Figure2}(b). The top right region is displaced by 1 pixel rightwards and downwards, mimicking 20.7nm of GBS. Three discrete Slip Bands (SBs) were added to analyze the effect of near-GB crystallographic slip. Additionally, noise was added to the images (representative of experimental conditions). The displaced and reference image are correlated using MatchID (version 2022.1), using DIC filtering settings motivated in Table 3 in \cite{vermeij_sslip_2025} (13nm pixelsize, 33pix\textsuperscript{2} subset size with affine shape functions, 3pix step size, ZNSSD algorithm, Gaussian prefiltering with 1pix filter size), leading to the effective strain $\hat{E}_{eff}$ and $\hat{\mathbf{H}}^{exp}_{2D}$ fields, shown in \autoref{fig:Figure1}(a,b\textsubscript{1-4}). Due to the DIC spatial filtering, discrete sliding/slip motions result in diffuse strain bands.

SLIDE is performed for each datapoint, yielding a field of sliding vectors \SvecHatTwo; their magnitude \GammaHatTwo\ is shown in \autoref{fig:Figure1}(c\textsubscript{1}). Results for $\hat{\theta}^{s}$ (the angle between \SvecHatTwo\ and \TvecBarTwo) and the normalized residual \RHat, are displayed in \autoref{fig:Figure1}(d,e).

To recover the discrete sliding displacement at the GB from the diffuse \SvecHatTwo\ field, the \SvecHatTwo\ vectors are integrated along \NvecBarTwo\ for each 1-pixel-long GB segment, see \autoref{fig:Figure1}(f). The integration path (solid magenta line), is aligned to \NvecBarTwo; the magenta vectors along the line are interpolated from the datapoint-wise \SvecHatTwo\ data and subsequently integrated, yielding the GB-integrated GBS vector \SvecBarTwo. The magnitude of this vector, \GammaBarTwo, is overlaid on the GB in \autoref{fig:Figure1}(f). The result for \GammaBarTwo\ along the entire GB length is plotted in red in  \autoref{fig:Figure1}(g), demonstrating that the in-plane displacement of 20.7nm is recovered, except for edge regions where full integration is not possible, and locations where SBs intersect the GB, marked in gray. \autoref{fig:Figure1}(g) also shows that \RBar\ (blue line) is high at GB segments where SBs intersect, because the measured displacement gradient tensor does not fit the kinematic description of GBS. By thresholding on \RBar, these unreliable regions are filtered out, a critical step that has been overlooked in previous works. \autoref{fig:Figure1}(h), showing the effect of the integration path length, reveals that due to DIC spatial filtering, a Minimal Integration Path Length (IPL\textsubscript{min}) of 700nm is required to accurately recover the discrete GBS, shown by the white arrow in \autoref{fig:Figure1}(f).

For the out-of-plane part of the virtual experiment, a calibration specimen with a known 19.4nm step is used, implying a magnitude of the 3D sliding vector \SvecBar\ of 28.4nm, see \autoref{fig:Figure1}(i). Using a Sensofar Sneox confocal OP (x150 magnification, "CSDS" algorithm), the specimen is measured. To yield the height step $d\bar{Z}$ at the GB for each 1-pixel-long GB segment, a 1500nm line is traced along \NvecBarTwo\ (see \autoref{fig:Figure1}(j)), of which the first and last 200nm are subtracted. The vertical displacement of 19.4nm is closely recovered, see \autoref{fig:Figure1}(k). Combining the in-plane sliding vector \SvecBarTwo\ and the height step $d\bar{Z}$, obtained with SLIDE, yields \SvecBar, whose magnitude \GammaBar\ matches the expected outcome, shown by the red line in \autoref{fig:Figure1}(l). Importantly, also the correct out-of-plane sliding angle $\bar{\beta}^{s}$ of 43$^\circ$ (blue line) is recovered.

After this successful virtual validation, the SLIDE framework is applied to a complicated real case of GBS/GBO as observed during an \textit{in-situ} SEM-DIC tensile test of a Zn-coated  (10$\mu m$ thick coating) DX54-galvanized steel. 20 Regions of Interest (ROIs) of $\sim$50x50$\mu m^2$ have been characterized and traced during deformation, one of which is highlighted in \autoref{fig:Figure2}. 

First, after polishing using a non-dry oxide polishing suspension, EBSD patterns (Edax Digiview 2) were indexed through spherical indexing using EMSphInx \cite{lenthe_spherical_2019}, see \autoref{fig:Figure2}(a). Subsequently, a high-quality InSn SEM-DIC speckle pattern was applied, resulting in 90-100nm sized speckles as  shown in \autoref{fig:Figure2}(b). The dogbone specimen (gauge cross section 4x0.7mm) was deformed \textit{in situ} in a Kammrath\&Weiss tensile stage in a Tescan Mira 3 SEM. The test was interrupted (with specimen unloaded) at five increments of deformation for imaging, see the stress-strain curve in \autoref{fig:Figure2}(c). The details of the imaging conditions and DIC patterning and filtering settings are given in \cite{vermeij_sslip_2025}. The alignment framework of Vermeij \textit{et al.} was employed, yielding $\sim$100nm accurate spatial alignment of EBSD and DIC data \cite{vermeij_nanomechanical_2022}. Ultimately, all data was aligned on grids with a pixelsize of 50nm. For analysis/plotting, the MTEX toolbox was used \cite{bachmann_texture_2010}. Note that recent SEM-DIC works often employ large-scale high-resolution stitched DIC data \cite{yavuzyegit_mapping_2023,linne_effect_2020,jullien_grain_2024,liu_-plane_2019,black_high-throughput_2023,chen_high-resolution_2018}, encompassing ROIs much larger than those presented here. The alignment framework of Vermeij \textit{et al.} (\cite{vermeij_nanomechanical_2022}) may be less accurate for large stitched datasets, whereas alternative methods might provide better results \cite{black_high-throughput_2023,linne_data_2019}. Regardless, alignment inaccuracies are resolved by tuning IPL\textsubscript{min} such that all GB deformation is accurately integrated over.

An example of an $\hat{E}_{eff}$ field is shown in \autoref{fig:Figure2}(d\textsubscript{1}). All analysis in this work is performed using this strain increment (global strain ${\varepsilon}_g = 5.88\%$). The results for all 5 increments of deformation are presented in more detail in the supplementary data. The DIC-EBSD alignment is visualized by the overlaid GBs in red (the same $\hat{E}_{eff}$ field without overlaid GBs is shown in \autoref{fig:Figure5}(a), more clearly visualizing the strain localization on the GBs). Note that the (GB) deformation can be severe, resulting in localized degradation of the DIC pattern, requiring interpolation of the displacement data \cite{vermeij_martensite_2023}, see \autoref{fig:Figure2}(d\textsubscript{2,3}).

The GBs are smoothened to accurately estimate \TvecBarTwo\ and \NvecBarTwo. The optimal amount of smoothing (using MTEX) can be calibrated for each dataset; an example of a sufficiently smooth GB is shown in \autoref{fig:Figure2}(e). All near-GB datapoints are assigned the nearest GB normal/tangential to perform SLIDE, see \autoref{fig:Figure2}(e). 

To distinguish between GBS and GBO, normalized IBSE images (also used for DIC) at the final increment of deformation are used, shown in \autoref{fig:Figure2}(f\textsubscript{1}). The GB data can be forward-deformed by performing either point alignment to the deformed IBSE image or DIC-based forward-deformation \cite{vermeij_nanomechanical_2022}. The close-ups in \autoref{fig:Figure2}(f\textsubscript{2,3}) show examples of GBS and GBO, and also show the alignment between forward-deformed GBs and the IBSE image in the deformed configuration. 

The out-of-plane displacements are obtained using OP (Sensofar Sneox, same settings), shown in \autoref{fig:Figure2}(g\textsubscript{1}), including forward-deformed GBs. \autoref{fig:Figure2}(g\textsubscript{2}) shows the absolute height gradient, showing peaks due to GBS, thereby also visualizing the alignment accuracy. In case of significant (etching-induced) initial height differences at the GBs, the out-of-plane displacement measurement accuracy is reduced, which can be restored by using the difference between the aligned height maps in undeformed and deformed configuriation \cite{liu_-plane_2019}.

\begin{figure}[ht!]
    \centering
    \includegraphics[width=\linewidth]{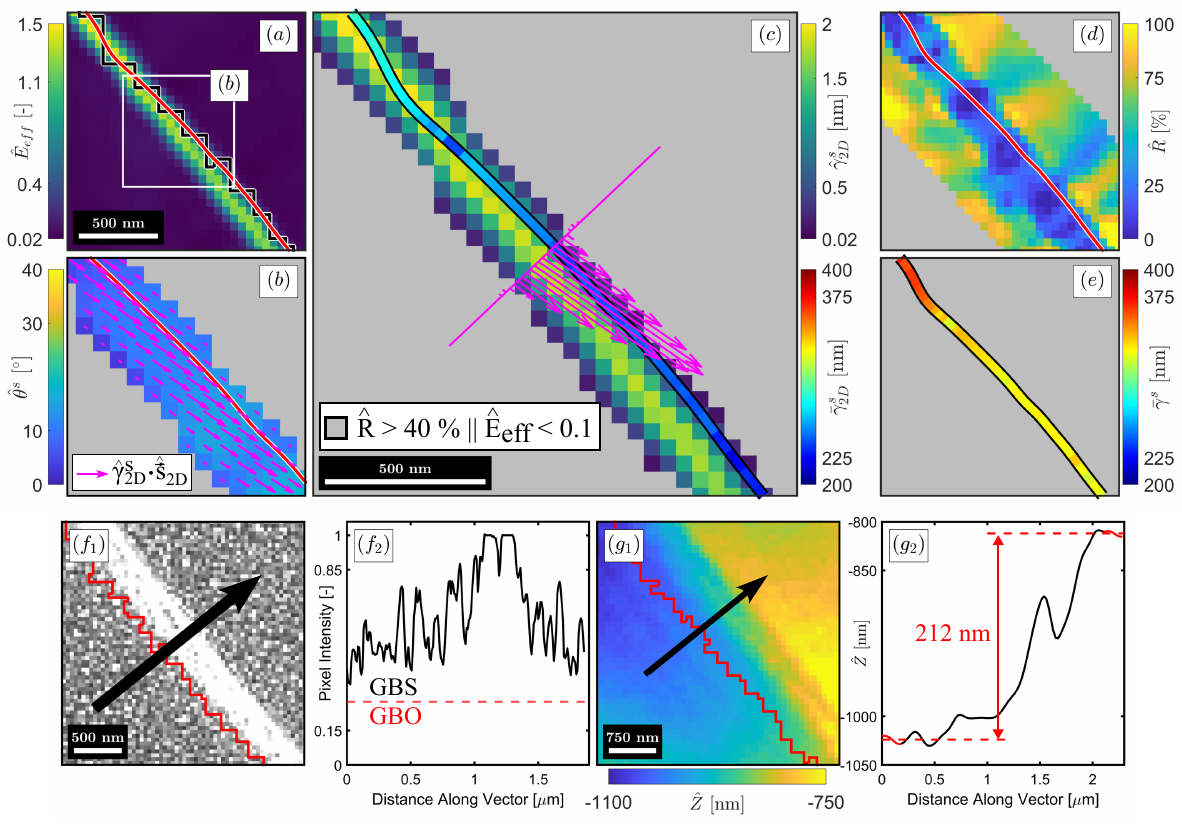}
    \caption{\textbf{Detailed example of quantification of grain boundary sliding.} (a) $\hat{E}_{eff}$ field, corresponding to the green region in \autoref{fig:Figure2}(d\textsubscript{1}), including raw (black) and smooth (red) GB. (b) Close-up showing in-plane misorientation angle $\hat{\theta}$, including smooth GB in red and \SvecHatTwo\ vectors in magenta. (c) \GammaHatTwo\ field, with \GammaBarTwo\ overlaid on the GB. An example of a 900nm integration line including interpolated \SvecHatTwo\ vectors is given in magenta. (d) Residual $\hat{R}$. The datapoints with $\hat{R}>40\%$ or $\hat{E}_{eff}<0.1$ are omitted from analysis and plotted in gray in (b,c). (e) 3D sliding magnitude \GammaBar\ overlaid on the smooth GB. (f\textsubscript{1}) IBSE data, including forward-deformed GB. For one $\sim$50nm GB segment, indicated by a black arrow, the average IBSE data along the GB normal is shown in (f\textsubscript{2}), including a red dashed line indicating the threshold for GBO. (g\textsubscript{1}) Profilometry data, including forward-deformed GB. For one GB segment indicated by a black arrow, the Z data along the GB normal is shown in (g\textsubscript{2}), and the height of the sliding step $d\bar{Z}$ is indicated (defined as the difference of the mean values of the first and last 200nm of the line segment, shown in red). }
    \label{fig:Figure3}
\end{figure} 

\begin{figure}[ht!]
    \centering
    \includegraphics[width=\linewidth]{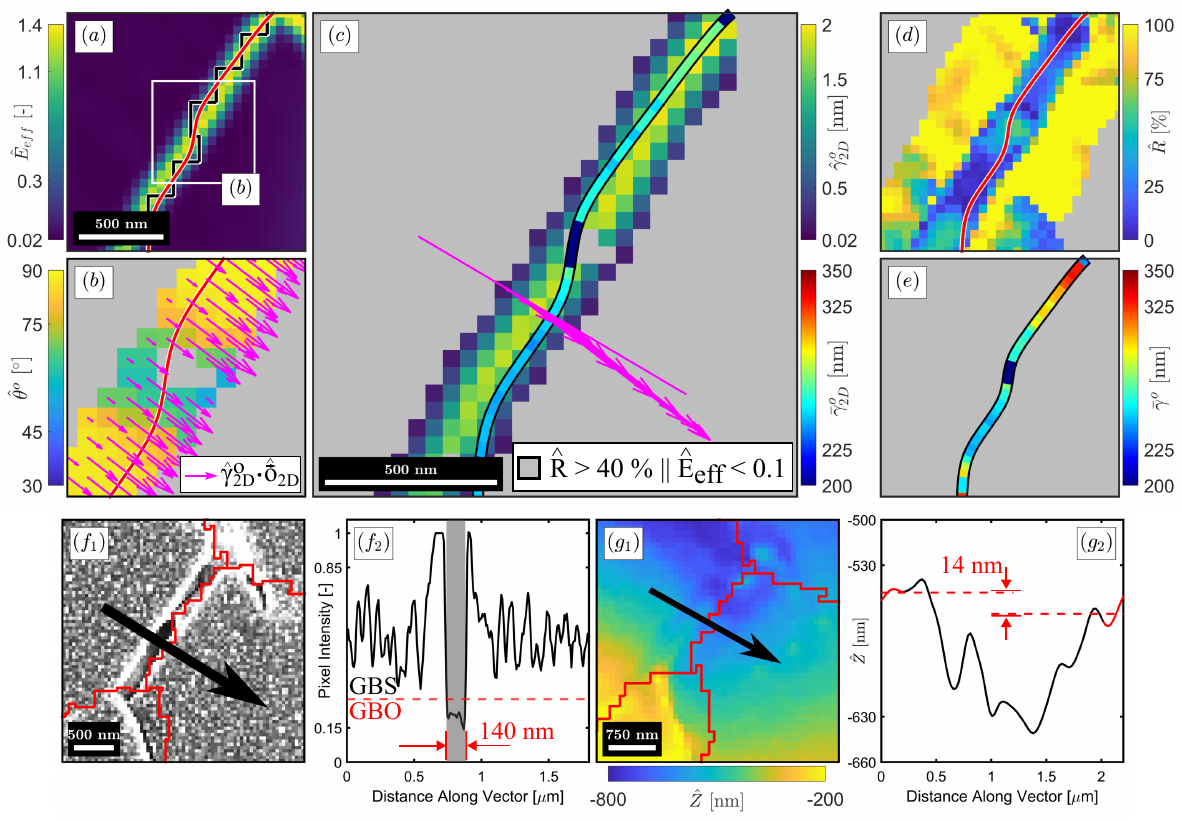}
    \caption{\textbf{Detailed example of quantification of grain boundary opening.} The location of this region corresponds to the white square in \autoref{fig:Figure2}(d\textsubscript{1}). For a complete description of the subfigures, see \autoref{fig:Figure3}. (f\textsubscript{2}) The average normalized IBSE data (brightness scaled between 0 and 1) along the single $\sim$50nm GB segment displayed by the black arrow in (f\textsubscript{1}) falls below the brightness threshold for GBO, set at 0.27. If the brightness falls below this threshold for 3 or more consecutive pixels (\textgreater 25nm), opening is identified. }
    \label{fig:Figure4}
\end{figure}

\begin{figure}[ht!]
    \centering
    \includegraphics[width=\linewidth]{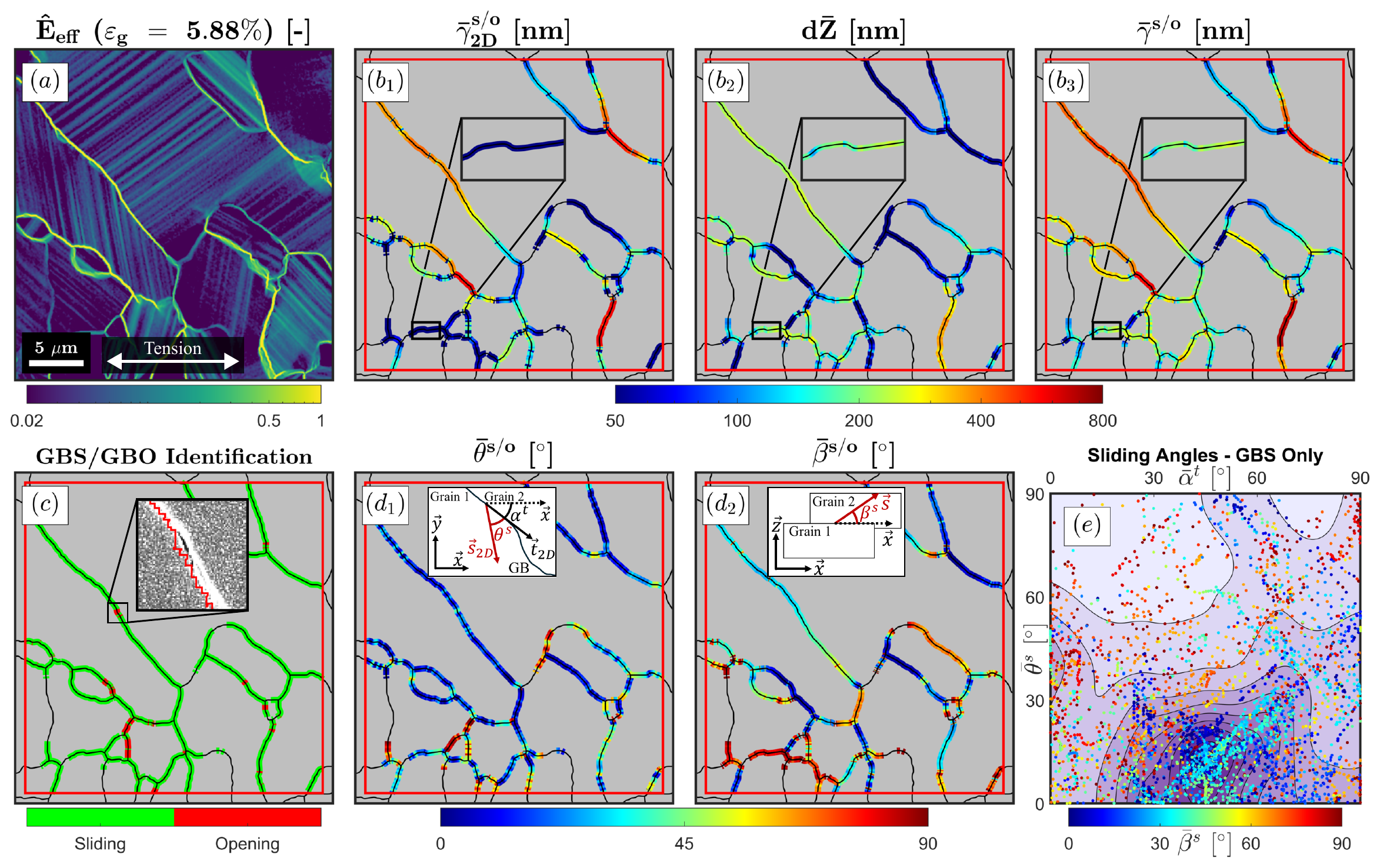}
    \caption{\textbf{Results of SLIDE, automatically applied to the complete region of \autoref{fig:Figure2}.} (a) Effective strain ($\hat{E}_{eff}$) map, in which the direction of tension is indicated. (b) In-plane and out-of-plane displacements recovered by SLIDE: (b\textsubscript{1}) magnitude of GB-integrated in-plane sliding or opening vector \GammaBarTwoCombi, (b\textsubscript{2}) height step $d\bar{Z}$, (b\textsubscript{3}) magnitude of complete sliding or opening vector \GammaBarCombi. (c) GBS/GBO identification based on IBSE data. (d\textsubscript{1}) Angle $\bar{\theta}^{s/o}$ between \SvecBarTwo/\OvecBarTwo\ and \TvecBarTwo. (d\textsubscript{2}) Angle $\bar{\beta}^{s/o}$ between $\vec{x}$ and \SvecBar/\OvecBar. (e) Correlation between $\bar{\theta}^{s}$ and $\bar{\alpha}^{t}$ (angle between \TvecBarTwo\ and $\vec{x}$) for all GB segments that show GBS. The colors indicate the out-of-plane angle $\bar{\beta}^{s}$, the background color represents the density of data.}
    \label{fig:Figure5}
\end{figure}

We demonstrate SLIDE in detail on two small regions (\autoref{fig:Figure3} and \autoref{fig:Figure4}) of the experimental dataset. \autoref{fig:Figure3} contains an example of pure GBS. The \SvecHatTwo\ vectors (\autoref{fig:Figure3}(b)) align closely to the GB tangential, making GBO in this region highly unlikely. The \GammaHatTwo\ field is shown in \autoref{fig:Figure3}(c), including the GB-integrated in-plane sliding displacement \GammaBarTwo. IPL\textsubscript{min} was increased from 700nm to 900nm to accommodate errors in the alignment procedure \cite{vermeij_nanomechanical_2022} and potential GB migration. \autoref{fig:Figure3}(d) shows low (blue) \RHat-values within the sliding band. The IBSE data in \autoref{fig:Figure3}(f\textsubscript{1,2}) displays a bright band, suggesting strong out-of-plane displacements. From OP, a height step of 212nm is extracted, see \autoref{fig:Figure3}(g\textsubscript{1,2}). Due to the large out-of-plane displacements, the 3D GBS magnitude \GammaBar\ in \autoref{fig:Figure3}(e) is significantly larger than the 2D sliding magnitude in \autoref{fig:Figure3}(c). A path length of 1800nm was used for the IBSE/profilometry data, to accommodate all possible inaccuracies in the forward-deformed alignment.

An example of GBO is shown in \autoref{fig:Figure4}. The \OvecHatTwo\ vectors resolved by SLIDE in \autoref{fig:Figure4}(b) are pointed almost perpendicular to the GB tangential \TvecBarTwo, resulting in high values for $\hat{\theta}^{o}$. Note that some datapoints could not be resolved by SLIDE due to local variations in \TvecBarTwo, which could be
resolved by more (local) smoothening, at the expense of a reduction of the precise GB location/angle. Nevertheless, a consistent solution is found for the surrounding datapoints. The high $\hat{\theta}^{o}$ values indicate likely GBO, but this cannot be confirmed by in-plane DIC data only, because a large misorientation between \SvecBarTwo/\OvecBarTwo\ and \TvecBarTwo\ could also be caused by strong out-of-plane GBS and/or a large out-of-plane GB tilt angle $\bar{\beta}$. The IBSE data in \autoref{fig:Figure4}(f\textsubscript{1}) and the corresponding line plot along \NvecBarTwo\ in \autoref{fig:Figure4}(f\textsubscript{2}), however, reveal a clear dark zone, indicative of GBO. The profilometry data of the same segment shows a valley, with a small $d\bar{Z}$ of 14nm. This confirms GBO, since a large in-plane $\hat{\theta}$ can only correspond to GBS when there is also a significant out-of-plane displacement, which is not observed here. 

The automatically generated SLIDE results for the full region from \autoref{fig:Figure2} are shown in \autoref{fig:Figure5}, where \GammaBar\ and \GammaBarO\ are shown in the same maps. The $\hat{E}_{eff}$ field in \autoref{fig:Figure5}(a) shows that (almost) all GBs are deforming; the combined GB strain concentrations are comparable to the combined intragranular slip. The combination of \GammaBarTwoCombi\ (\autoref{fig:Figure5}(b\textsubscript{1})) and $d\bar{Z}$ (\autoref{fig:Figure5}(b\textsubscript{2})) yields the 3D sliding/opening vectors \SvecBar\ and \OvecBar\ whose magnitudes are shown in \autoref{fig:Figure5}(b\textsubscript{3}). Note the zoomed region in \autoref{fig:Figure5}(b\textsubscript{1-3}), which has a small in-plane contribution but significant out-of-plane displacements, illustrating the necessity of 3D sliding identification. \autoref{fig:Figure5}(c) shows the occurrence of GBS/GBO. The close-up shows IBSE data of a small GB segment, almost vertically oriented, which opens, although the rest of the more diagonally oriented GB slides. This illustrates that the local orientation of the GB has a strong influence on its sliding/opening behavior, highlighting the importance of local nanoscale GBS/GBO identification. 

The angles $\bar{\theta}^{s/o}$ and $\bar{\beta}^{s/o}$ are shown in \autoref{fig:Figure5}(d\textsubscript{1,2}). \autoref{fig:Figure5}(e) shows the correlation between $\bar{\alpha}^{t}$, $\bar{\beta}^{s}$ and $\bar{\theta}^{s}$ (for GBS only). GBS is mostly observed for GBs oriented $\sim$45$^\circ$ from the direction of tension; these sliding events are closely aligned to the GB tangential (low $\bar{\theta}^{s}$) and show moderate out-of-plane activity. However, for GBs oriented horizontally/vertically in-plane (low/high $\bar{\alpha}^{t}$), stronger out-of-plane activity (larger $\bar{\beta}^{s}$) is observed. A similar plot is available for GBO (not shown), equally rich in relevant data. These and other statistical GB analyses, unlocked by SLIDE, may well reveal new micromechanical insights or provide guidelines for GB/texture engineering to optimize grades. Finally, the nature of the local interactions of GBS/GBO with intragranular slip in the neighbouring grains is of great interest \cite{ando_internal_2016,linne_effect_2020,jullien_grain_2024}, for which SLIDE and SSLIP \cite{vermeij_automated_2023} analyses can be readily combined. For example, GBO can be correlated to intragranular slip (incompatibility) in the neighbouring grains, gaining a deeper understanding of the opening mechanisms of the Zn coating, potentially providing guidelines to improve its resistance to damage. This analysis, performed for all 20 ROIs monitored during the here-presented \textit{in-situ} test, reveals the critical GB configurations responsible for GBO, as will be presented in a future publication. Such statistical GBS/GBO studies open opportunities for direct comparison with TEM-based observations of atomic-scale GB deformation mechanisms \cite{wang_tracking_2022,rajabzadeh_role_2014}.

In summary, we propose a novel framework to accurately identify and quantify 3D GBS/GBO based on precisely aligned DIC, EBSD, IBSE, and OP data, named \textbf{S}liding identification by \textbf{L}ocal \textbf{I}ntegration of \textbf{D}isplacements across \textbf{E}dges (\textbf{SLIDE}). SLIDE has been validated on virtual and experimental data, demonstrating that 3D GBS/GBO is locally distinguished and quantified while non-GB plasticity/elasticity/noise is rejected, improving upon previous works. The combination of accurate nanoscale GBS/GBO quantification and fast analysis of complete experimental datasets enables the analysis of GB mechanics on a larger scale, which opens avenues to fully understanding of the intricate deformation behavior of GBs in complex alloys.

\subsection*{Code and Data Availability}

The Matlab/MTEX code for the SLIDE framework, including an example using the experimental Zn coating data discussed in this manuscript, is available on Github: \url{https://github.com/CasperMornout/SLIDE}.

\subsection*{Acknowledgements}

We acknowledge Marc van Maris and Mark Vissers for experimental support. This research was carried out as part of the “Next-Coat” and "MarBend" projects, under project numbers N19016b and T22005, in the framework of the Partnership Program of the Materials innovation institute M2i (\url{www.m2i.nl}) and the Netherlands Organization for Scientific Research (\url{http:// www.nwo.nl}). 

\subsection*{CRediT Author Statement}
\textbf{C.J.A. Mornout}: Methodology, Software, Validation, Formal Analysis, Investigation, Writing - Original Draft, Visualization, Data Curation.
\textbf{G. Slokker}: Methodology, Software, Investigation, Writing - Review \& Editing.
\textbf{T. Vermeij}: Conceptualization, Software, Methodology, Writing - Review \& Editing, Supervision.
\textbf{D. König}: Conceptualization, Writing - Review \& Editing, Supervision.
\textbf{J.P.M. Hoefnagels}: Conceptualization, Methodology, Resources, Writing - Review \& Editing, Supervision, Project administration, Funding acquisition.

\bibliographystyle{unsrt}


\end{document}